\documentclass[pra,twocolumn,showpacs,groupedaddress,superscriptaddress,aps,10pt]{revtex4-1}

\usepackage{bm,graphicx,amsmath}
\usepackage{placeins}
\usepackage{amssymb}
\usepackage{amsmath}
\usepackage{epsfig}
\usepackage{amssymb}
\usepackage{color}
\usepackage[pdftex,colorlinks,linkcolor=blue,citecolor=blue]{hyperref}
\usepackage{subfigure}
\newcommand{\ket}[1]{\ensuremath{\left| #1 \right\rangle}}

\newcommand{\1}{\ensuremath{\left|1\right\rangle}}
\newcommand{\2}{\ensuremath{\left|2\right\rangle}}
\newcommand{\3}{\ensuremath{\left|3\right\rangle}}

\newcommand{\bmu}{\ensuremath{\bm{\mu}}}
\newcommand{\be}{\ensuremath{\bm{E}}}

\begin{document}

\title{Nanoscale resolution for fluorescence microscopy via adiabatic passage}

\author{J. L. Rubio}
\affiliation{Departament de F\'{\i}sica, Universitat Aut\`{o}noma de Barcelona, E-08193 Bellaterra, Spain} 
\author{D. Viscor}
\affiliation{Departament de F\'{\i}sica, Universitat Aut\`{o}noma de Barcelona, E-08193 Bellaterra, Spain}
\author{V. Ahufinger}
\affiliation{Departament de F\'{\i}sica, Universitat Aut\`{o}noma de Barcelona, E-08193 Bellaterra, Spain}
\author{J. Mompart}
\affiliation{Departament de F\'{\i}sica, Universitat Aut\`{o}noma de Barcelona, E-08193 Bellaterra, Spain}

\begin{abstract}
We propose the use of the subwavelength localization via adiabatic passage technique for fluorescence microscopy with nanoscale resolution in the far field. 
This technique uses a $\Lambda$-type medium coherently coupled to two laser pulses: the pump, with a node in its spatial profile, and the Stokes. The population of the $\Lambda$ system is adiabatically transferred from one ground state to the other except at the node position, yielding a narrow population peak.
This coherent localization allows fluorescence imaging with nanometer lateral resolution.
We derive an analytical expression to asses the resolution and perform a comparison with the coherent population trapping and the stimulated-emission-depletion techniques.
\end{abstract}

\maketitle

\section{Introduction}
In the last decades, far-field fluorescence microscopy has experienced great advances with applications in medicine and biology, among other fields, offering the possibility to obtain high resolution images in a non-invasive manner. In lens-based light microscopes, the image of a point object obtained from the fluorescence emitted by a fluorophore placed in the sample is, in principle, limited by diffraction \cite{Abbe'73}.
This image becomes a finite-size spot in the focal plane, mathematically described by the point-spread function (PSF), whose full width at half maximum (FWHM) is $\lambda/(2\,\rm NA)$, being $\lambda$ the wavelength of the addressing light, and $\rm NA$ the numerical aperture of the objective. Due to the reduced dimensions of the samples to investigate, frequently around few nanometers, high resolution images overcoming the diffraction limit are necessary.
To this aim, one of the most extended group of techniques is based on the general concept of reversible saturable optical fluorescence transition (RESOLFT) \cite{Hell'03} between two distinguishable molecular states, such as stimulated-emission-depletion (STED) \cite{Hell'94} and ground-state-depletion (GSD) \cite{Hell'95}. RESOLFT exploits the spatial inhibition of the fluorescence signal from a light-excited fluorophore in order to engineer the effective PSF, reducing its FWHM.

Parallely, in recent years, there has been an intense activity in the spatial localization of atomic population in $\Lambda$-type systems coherently interacting with two electromagnetic fields (see e.g., \cite{Agarwal'06,Gorshkov'08}). All these methods are based on the so-called coherent population trapping (CPT) technique or close variations, and have also been considered for microscopy \cite{Yavuz'07, Kapale'10,Li'08}).
In the context of atomic population localization, some of us have recently proposed the subwavelength localization via adiabatic passage (SLAP) technique~\cite{Mom'09}.
In SLAP, position-dependent stimulated Raman adiabatic passage (STIRAP) \cite{Bergmann'98} is performed using spatially dependent fields. This approach provides population peaks narrower than using other coherent localization techniques \cite{Agarwal'06,Yavuz'07} and its application in nanolitography and patterning of BEC's \cite{Mom'09}, and single-site addressing of ultracold atoms \cite{Vis'12}, has already been discussed.

In this work, we propose a SLAP-based nanoscale resolution technique for fluorescence microscopy.
At variance with respect to \cite{Mom'09, Vis'12}, two driving laser pulses are applied here to a continuous distribution of emitters, and and we take into account the effect of the objective lens in our model.
In section \ref{sec:Model}, we describe the physical system under consideration, derive an expression for the resolution, and compare it with the one obtained for the CPT case. In section \ref{sec:NumericalResults}, we show numerical results of the SLAP proposal and compare it with the STED technique. Finally, we present the conclusions and point out a possible implementation of the proposed technique using quantum dots.

\section{Model}
\label{sec:Model}

\begin{figure}

\includegraphics[width=0.8\columnwidth]{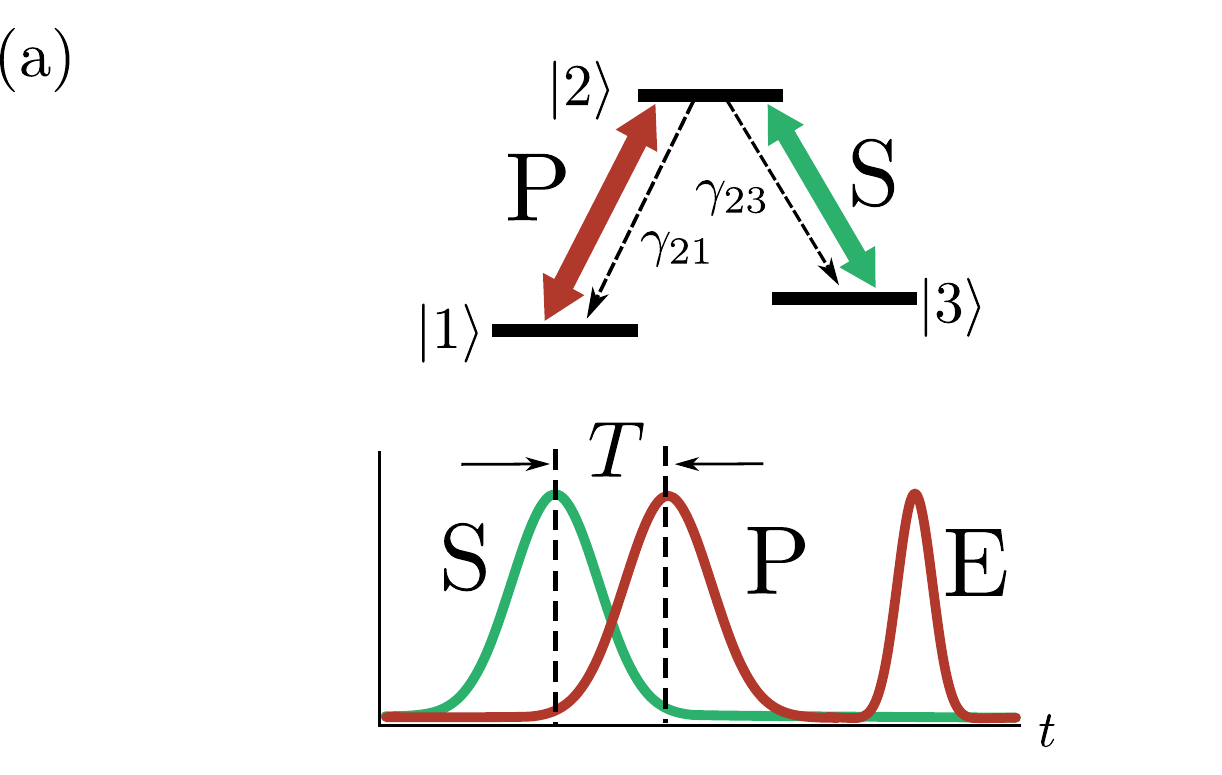}
\includegraphics[width=0.8\columnwidth]{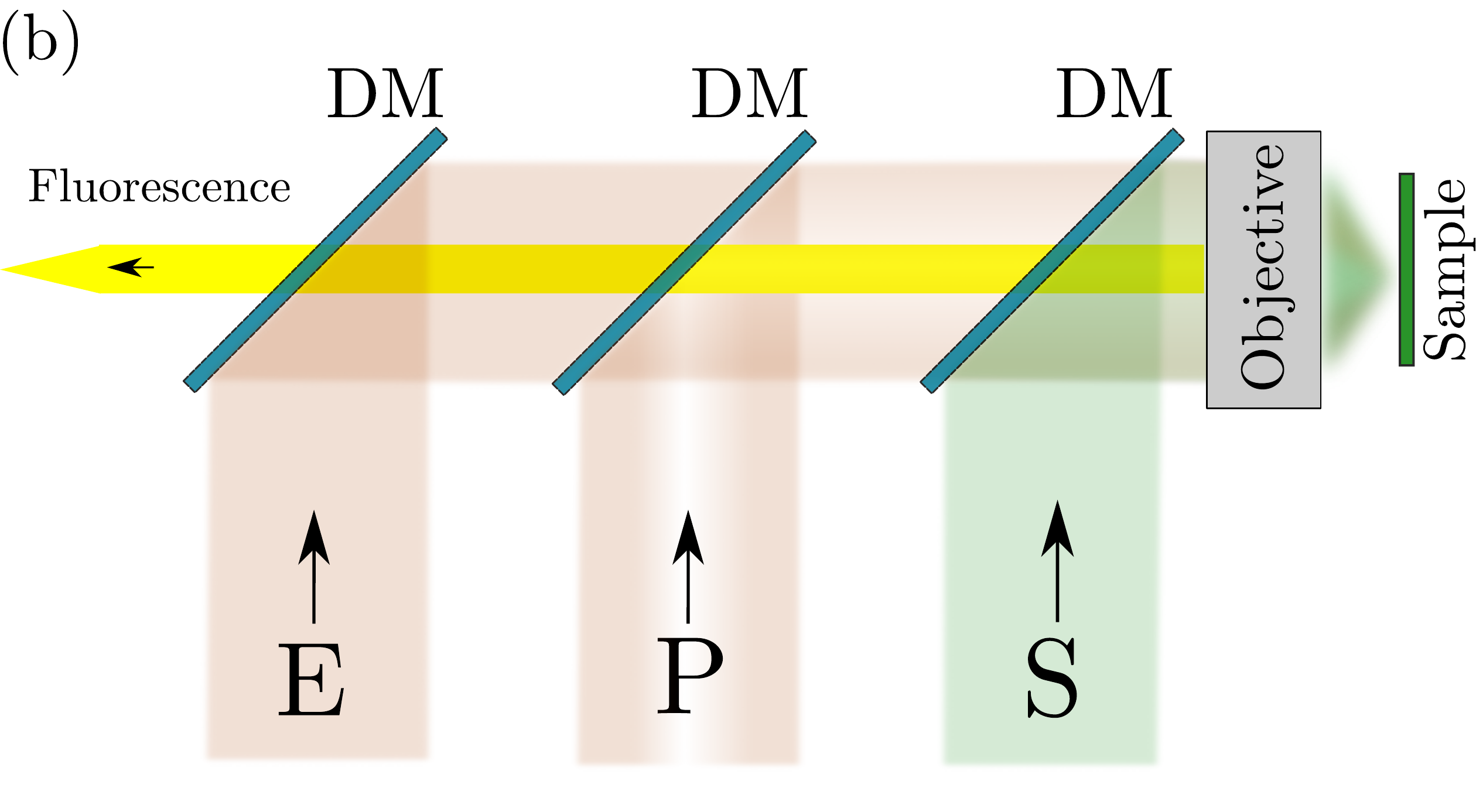}
\caption{
(a) SLAP technique. Top: $\Lambda$-system with pump (P) and Stokes (S) pulses coupling the \1$\leftrightarrow$\2 and \3$\leftrightarrow$\2 transitions, respectively.
The decay rate from \2 to \1 (\3) is $\gamma_{21}$ ($\gamma_{23}$). Bottom: Pulse temporal sequence, being E the exciting pulse. (b) Schematic setup of the SLAP-based fluorescence microscopy. Note the central node of the pump pulse. DM accounts for dichroic mirrors.}
\label{f:fig1}
\end{figure}

We consider a $\Lambda$ scheme, where the population is initially in \1, see top of Fig. \ref{f:fig1}(a).
Two laser pulses, Stokes (S) and pump (P), couple to transitions \3$\leftrightarrow$\2 and \1$\leftrightarrow$\2 with Rabi frequencies $\Omega_S(x,t)\equiv \bmu_{32} \cdot{\be_{\rm S}}(x,t)/\hbar$ and $\Omega_P(x,t)\equiv\bmu_{12}\cdot{\be_{\rm P}}(x,t)/\hbar$, respectively, where ${\be_{\rm S}}$ (${\be_{\rm P}}$) is the electric field amplitude of the Stokes (pump), $\bmu_{32}$ ($\bmu_{12}$) is the electric dipole moment of the \3$\leftrightarrow$\2 (\1$\leftrightarrow$\2) transition, and $\hbar$ is the reduced Planck constant. The excited level \2 has a decay rate $\gamma_{21}$ ($\gamma_{23}$) to the ground state \1 (\3). If the two-photon resonance condition is fulfilled, one of the eigenstates of the Hamiltonian, the so-called dark-state, takes the form:
\begin{align}\label{eq:darkstate}
\ket{D(x,t)}=[\Omega_{\rm S}^{*}(x,t)\ket{1}-\Omega_{\rm P}^{*}(x,t)\ket{3}]/\Omega(x,t),
\end{align}
where $\Omega(x,t)=(|\Omega_{\rm P}(x,t)|^2+|\Omega_{\rm S}(x,t)|^2)^{1/2}$.
Note that $\ket{D(x,t)}$ does not involve the excited state \2.
In our scheme, and in order to adiabatically follow the dark state, both pulses are sent in a counterintuitive temporal sequence, applying first the Stokes and, with a temporal delay $T$, the pump [see bottom of Fig. \ref{f:fig1}(a)]. In addition, to ensure no coupling between the different energy eigenstates, a global adiabatic condition must be fulfilled, which imposes $\Omega(x)\,T\geq A$, where $A$ is a dimensionless constant that for optimal Gaussian profiles and delay times takes values around 10 \cite{Bergmann'98}. In the SLAP technique, the pump pulse has a central node in its spatial profile such that the described temporal sequence of the pulses produces an adiabatic population transfer from \1 to \3, except at the position of the pump node. Therefore, we obtain a narrow peak of population remaining in \1, whose profile can be considered as the PSF function referred to an image object in a lens-based microscope. Note that the FWHM of the population peak determines the resolution of the technique. Finally, an exciting E pulse is used to pump the population remaining in \1 either to state \2 or to an auxiliary excited state, allowing for the subsequent registration of the fluorescence.
If state \2 decays radiatively, the pump (P) could be used also as the exciting (E) pulse, reducing the experimental requirements. Figure \ref{f:fig1}(b) shows a possible setup for the proposal. 
Right to left, the Stokes, pump and exciting pulses are sent, and we assume that due to the Stokes's shift, the fluorescence (left arrow) can be separated from the light of the different pulses by, e.g., dichroic mirrors (DM).


\begin{figure}[htbp]
\centering\includegraphics[width=1\columnwidth]{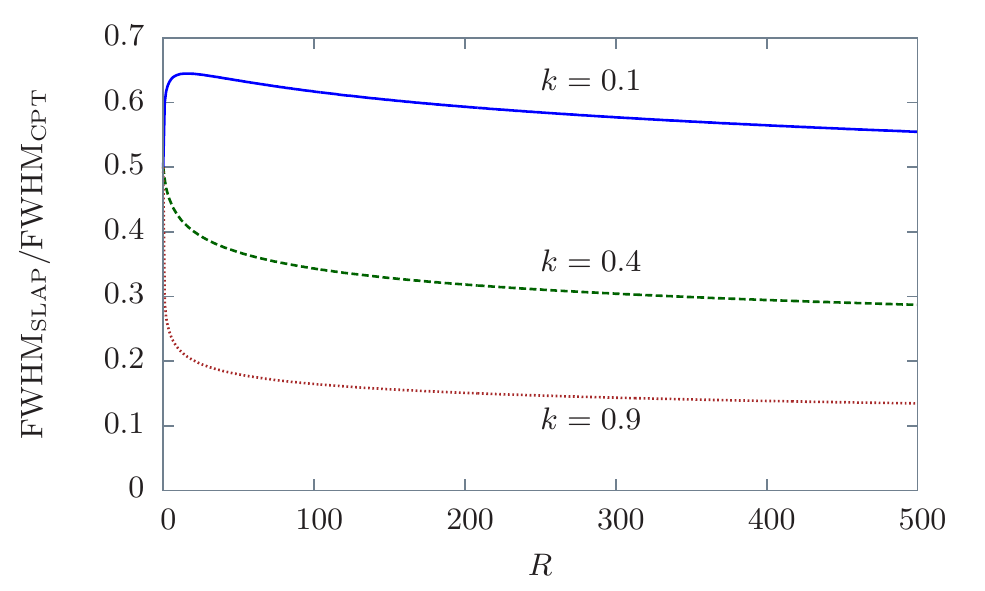}
\caption{
Ratio between the FWHM using SLAP and CPT techniques as a function of $R$, for $k=0.1$ (blue solid line), $k=0.4$ (green dashed line), and $k=0.9$ (red dotted line).}
\label{f:fig2}
\end{figure}

To take into account the effect of the objective lens in our model, we consider the Rabi frequency of the Stokes pulse having a Bessel beam spatial profile, and the one of the pump being the result of superimposing two Bessel beams focused with a lateral offset, producing a node in its center. The spatio-temporal profiles of the Stokes and pump fields read:
\begin{align}
\Omega_{\rm S}(\upsilon,t)&=\Omega_{\rm S0}F(\upsilon,0)\,e^{-(t-t_{\rm S})^2/\sigma^2}, \label{eq:fields1} \\ 
\Omega_{\rm P}(\upsilon,t)&=\Omega_{\rm P0}[F(\upsilon,\delta)+F(\upsilon,-\delta)]\,e^{-(t-t_{\rm P})^2/\sigma^2},\label{eq:fields2}
\end{align}
where $\Omega_{\rm S0}$ ($\Omega_{\rm P0}$) is the peak Rabi frequency of the Stokes (pump) pulse, $F(\upsilon,r)\equiv\frac{2J_{1}(\upsilon + r)}{\upsilon+r}$, where $J_{1}(\upsilon)$ is the first order Bessel function and $\upsilon=(2\pi x \rm NA)/\lambda$ is the optical unit corresponding to the Cartesian coordinate $x$ in the focal plane, $\sigma$ is the temporal width of the pulses, $T=t_{\rm P}-t_{\rm S}$ is the temporal delay between the pulses, which is proporcional to $\sigma$,
and $\delta=1.22\pi$ is the offset with respect to $\upsilon=0$ corresponding to the first cutoff of $F(\upsilon,0)$.
It is possible to obtain an analytical expression for the FWHM of the final population peak in \1, $p_1(x)$, by considering that the global adiabaticity condition is reached for $\left|\upsilon\right|\approx\rm FWHM$, assuming a Gaussian population peak profile and considering $\left|\upsilon\right|\ll\delta$. Thus, from the spatial profiles in Eqs. (\ref{eq:fields1})-(\ref{eq:fields2}), the FWHM of the population distribution is
\begin{align}\label{eq:FWHM-SLAP}
{\rm FWHM_{SLAP}}=\frac{\lambda}{2\rm NA}\frac{\delta}{\pi}\left(\sqrt{\frac{4R}{k^{-2}-1}}+1\right)^{-1/2},
\end{align}
where $R\equiv(\Omega_{\rm P0}/\Omega_{\rm S0})^2$ is the intensity ratio between the pump and the Stokes pulses, and $k\equiv\Omega_{\rm S0}T/A$ must fulfill $0<k<1$.
In such a SLAP-based fluorescence microscope, the effective PSF is given by the product
$h_{\rm exc}(\upsilon)\,p_1(\upsilon)$ normalized to 1, where $h_{\rm exc}(\upsilon)$ is the PSF of the E pulse. If we assume that all the localized population leads to fluorescence, the lateral resolution is determined by Eq. (\ref{eq:FWHM-SLAP}), where the first factor accounts for diffraction, the second one is 1.22, and the third one can be less than 1 depending on the adiabaticity of the process, and tends to zero in the limit $R\rightarrow\infty$, i.e., when $\Omega_{\rm P0}\gg\Omega_{\rm S0}$.

Other dark-state techniques proposed to obtain atomic localization \cite{Agarwal'06,Yavuz'07} are based on the so-called coherent population trapping (CPT) or close variations, in which the fields can be either two continuos waves or two perfectly overlapping long pulses. In this case, to obtain the analytical expression for the FWHM of the final population peak in state \1, we have considered, as in~\cite{Agarwal'06}, that $|\left\langle 1|D(\upsilon) \right\rangle|^2=1/2$ for $\upsilon=\rm FWHM/2$ .
Using the profiles given in (\ref{eq:fields1}) and (\ref{eq:fields2}), the FWHM for CPT is
\begin{align}\label{eq:FWHM-CPT}
{\rm FWHM_{CPT}}=\frac{\lambda}{2\rm NA}\frac{2\delta}{\pi}\left(\sqrt{2\sqrt{R}+1}\right)^{-1/2}.
\end{align}

%
Figure~\ref{f:fig2} shows the ratio between the analytical FWHM obtained for SLAP [Eq. (\ref{eq:FWHM-SLAP})] and CPT [Eq. (\ref{eq:FWHM-CPT})] as a function of $R$ for different values of $k$. From the figure, we can see that in the whole range of parameters considered, the peak obtained with SLAP is significantly narrower than the one with CPT. Note that the adiabatic nature of the SLAP technique allows to increase the final resolution by increasing the time delay $T$, while fixing the intensities.
\section{Numerical results}
\label{sec:NumericalResults}

\begin{figure}
{
\includegraphics[width=0.6\columnwidth]{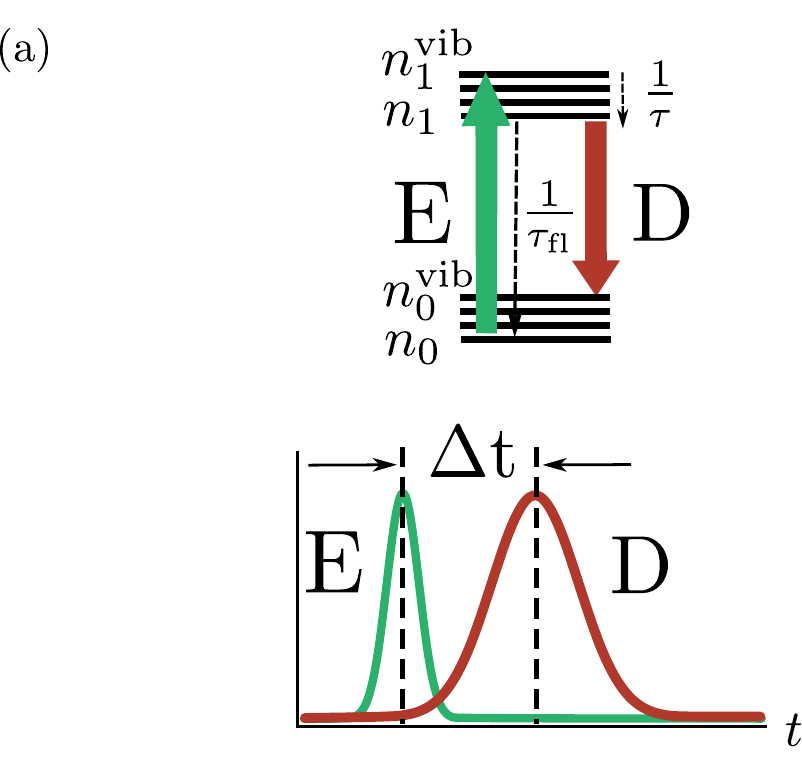}
\includegraphics[width=1\columnwidth]{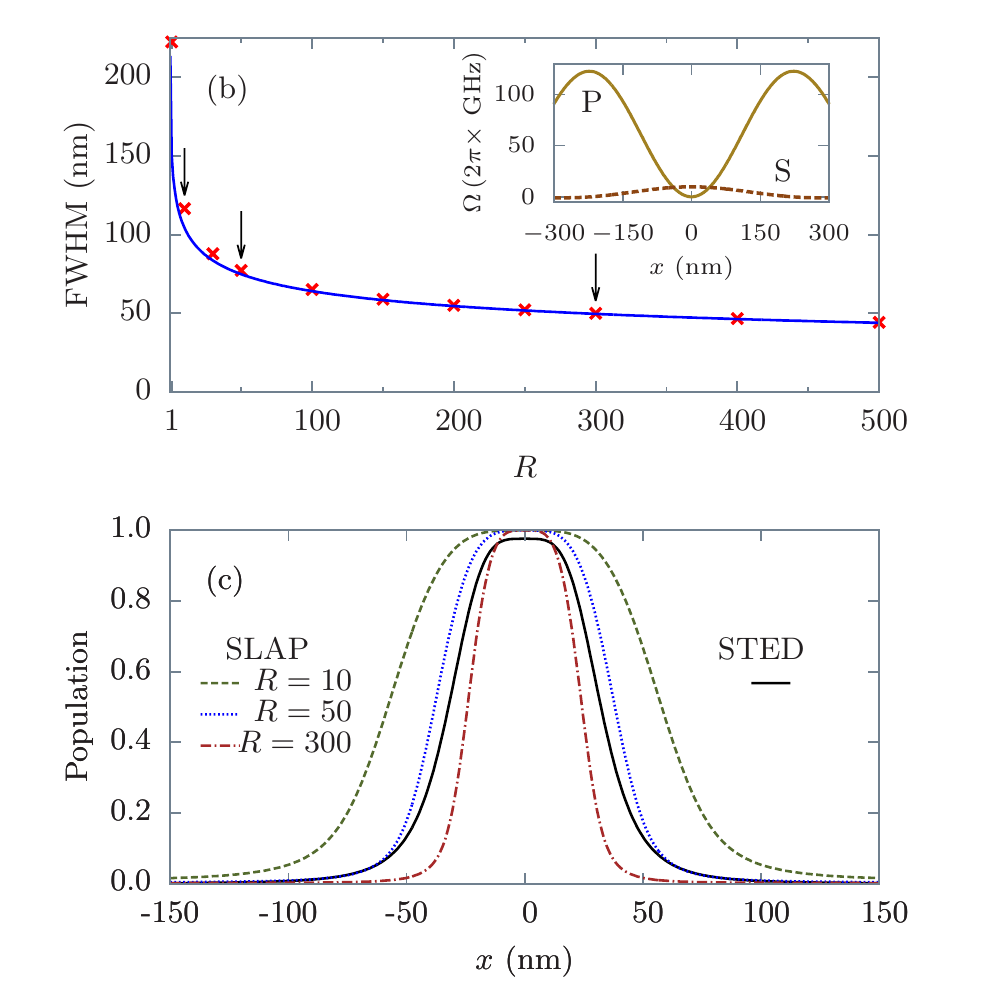}
}
\caption{(a) STED technique. Top: Fluorophore energy levels interacting with the exciting (E) and the depletion (D) fields, and their decay rates. Bottom: Pulse temporal sequence. (b) Analytical (solid line) and numerical (crosses) values for the FWHM of the population peak in \1, using SLAP, as a function of $R$. Inset: Spatial profiles of pump (P) and Stokes (S) Rabi frequencies for $R=150$. (c) Numerical results of the population $p_1(x)$ in \1 for SLAP using different values of $R$, and the population $n_1(x)$ in the first excited vibrational level of Rhodamine B for STED typical values (see text).}
\label{f:fig3}
\end{figure}


In the following, we are interested in comparing our scheme with the STED microscopy technique {[see Fig.~\ref{f:fig3}(a)]. In the STED technique, two beams, the exciting (E) and the depletion (D) lasers, interact with an organic fluorophore, with a time delay $\Delta t$. Typical values for $\Delta t$ are some hundreds of ps, larger than the vibrational relaxation time $\tau$ ($\sim$ps) but much shorter than the fluorescence time $\tau_{\rm fl}$ ($\sim$ns) of the transition $n_1\rightarrow n_0$. First, the E field excites all the population to state $n^{\rm vib}_1$, which rapidly decays to $n_1$. Next, the D field, which has a doughnut-like spatial profile, produces a spatial depletion of the population in $n_1$ by stimulated emission. Out of the node, the excited population is removed resulting in fluorescence inhibition, and reducing the width of the effective PSF. Note here that the main distinctive feature of SLAP with respect to general RESOLFT techniques, e.g., STED, is the adiabatic nature of the state transfer process, which as discussed below, confers robustness and flexibility on our method.

%
In Fig.~\ref{f:fig3}(b), values for the FWHM of the final population peak obtained from numerical simulations using SLAP (crosses) and the analytical curve (solid line) given by Eq. (\ref{eq:FWHM-SLAP}) using $A=20$ are represented.
For the simulations, we have used the density-matrix formalism for a $\Lambda$-system with degenerated ground states with the following parameter setting: $\gamma_{21}=\gamma_{23}=2\pi\times 6.36$ GHz, $\Omega_{\rm S0}/\gamma_{21}=1.5$, $\sigma=100$ ps, $T=1.5\sigma$, $\rm NA=1.4$, and $\lambda=490$ nm.
%
Figure~\ref{f:fig3}(c) shows the final population $p_1(x)$ in \1 using the SLAP technique for $R=10$ (dashed line), $R=50$ (dotted line) and $R=300$ (dotted-dashed line), marked with vertical arrows in Fig.~\ref{f:fig3}(b). In addition, the population $n_1(x)$ using the STED technique (black solid line) is also shown. For the STED simulation, we have used rate equations for the Rhodamine B dye with intensity profiles of the exciting and depletion pulses corresponding to Eq.~(\ref{eq:fields1}) and Eq.~(\ref{eq:fields2}), respectively, and typical values \cite{Hell'94} for the absorption cross sections $\sigma_{\rm cs}=10^{-17}$ cm$^2$, peak intensity of the depletion laser $h_{\rm D}^{\rm peak}=1300$ MW/cm$^2$, $\tau=1$ ps, $\tau_{\rm fl}=2$ ns, $\Delta$t $=90$ ps, $\sigma=100$ ps, $\rm NA=1.4$, $\lambda_{\rm E}=490$ nm, and $\lambda_{\rm D}=600$ nm, obtaining $\rm FWHM=65.2$ nm.
Note that the final peak in STED does not reach unity due to the loss of population by fluorescence while depletion acts. This does not occur in SLAP, since the final peak corresponds to the population in a ground state.
%

\section{Conclusions and perspectives}

In conclusion, we have presented a proposal to implement nanoscale resolution microscopy using the SLAP technique. 
We have derived an analytical expression to estimate the lateral resolution and we have compared it with the corresponding one for the CPT case, showing that SLAP yields a better resolution.
In both cases, the resolution can be improved by increasing the pump intensity $\propto\Omega_{\rm P0}^2$, provided $\Omega_{\rm S0}$ is kept constant. This behavior is similar in STED microscopy, whose resolution improves by increasing the intensity of the depletion laser.
Then, we have performed a numerical comparison between the STED technique using typical parameter values and the SLAP technique with Rabi frequencies of the order of GHz, obtaining a similar resolution in both cases.

All the previous results suggest that the localization via adiabatic passage may offer interesting features for fluorescence microscopy, for which coherent interaction between a $\Lambda$-type system and pump and Stokes fields would be required. In this sense, fluorescent semiconductor nano-crystals, also known as quantum-dots \cite{Alivisatos'96} (q-dots), could be good candidates to act as a diluted medium over the sample. Q-dots have excellent photostability, exhibit efficient fluorescence, can be selectively attached onto the item to study, and have been used effectively for imaging cells and tissues \cite{Michalet'05}. Further, coherent population transfer via adiabatic passage in two \cite{Hohenester'00} and three \cite{Fabian'05} coupled q-dots has been recently proposed. Q-dots have recombination times of the order of some hundreds of ps, corresponding to Rabi frequencies of GHz, and intensity of the pump beam around $10^6$ W/cm$^2$ for $\left|\bmu\right|\simeq10^{-28}$ C m \cite{Eliseev'00}. This intensity is similar to that used recently in RESOLFT-based microscopy using q-dots \cite{Irvine'07}, and up to three orders of magnitude below the one of the depletion beam used in STED microscopy with conventional fluorophores.
Nevertheless, due to the adiabatic nature of SLAP: (i) it is possible to enhance the resolution by increasing the temporal duration of the pulses, without increasing the fields intensity, thus reducing the possibility of damaging the sample, (ii) the spontaneous decay rate from \2 does not play any role in the localization process, and (iii) there is no need of using resonant lasers in our localization method provided the two photon resonance condition is fulfilled. This fact should permit to use the same  experimental arrangement with different types of emitters.



\section*{Acknowledgments}

We acknowledge David Artigas for fruitful comments, and funding from the Spanish Ministry of Economy and Competitiveness under Contract No. FIS2011-23719, and from the Catalan Government under Contract No. SGR2009-00347.

\end{document}